\documentstyle[aps,epsf,psfig]{revtex}

\newcommand{\HRule}{\rule{\linewidth}{1mm}}
\begin{document}
   \HRule
     \begin{center}
         \Huge {\bf ECT* PREPRINT}
     \end{center}
   \HRule
   \vspace*{\stretch{1}}
   \begin{center}
      \Large {\bf Self--energy of $\Lambda$ in finite nuclei}
   \end{center}
   \begin{center}
      \large M.\ Hjorth--Jensen$^a$, A.\ Polls$^b$,
       A.\ Ramos$^b$ and H.\ M\"uther$^c$
    \end{center}
    \begin{center}
      \large   $^a$ECT*, European Centre for Theoretical
        Studies in Nuclear Physics and Related Areas,
        Trento, Italy
    \end{center}
    \begin{center}
     \large $^c$Institut f\"ur Theoretische Physik, Universit\"at T\"ubingen,
     T\"ubingen, Germany
    \end{center}
    \begin{center}
      \large   $^b$ Departament d'Estructura i Constituents de la Materia,
       Universitat de Barcelona, Barcelona, Spain
    \end{center}
     \begin{center}
      \large $^c$Institut f\"ur Theoretische Physik, Universit\"at T\"ubingen,
        T\"ubingen, Germany
    \end{center}
    \vspace*{\stretch{2}}
    \begin{center}
      \large   Submitted to:  Nuclear Physics {\bf A}
    \end{center}
    \vspace*{\stretch{4}}
    \begin{center}
        \Large {\bf ECT* preprint $\#$: ECT*--96--008}
    \end{center}
    \begin{figure}[hbtp]
        \begin{center}
        {\centering\mbox{\psfig{figure=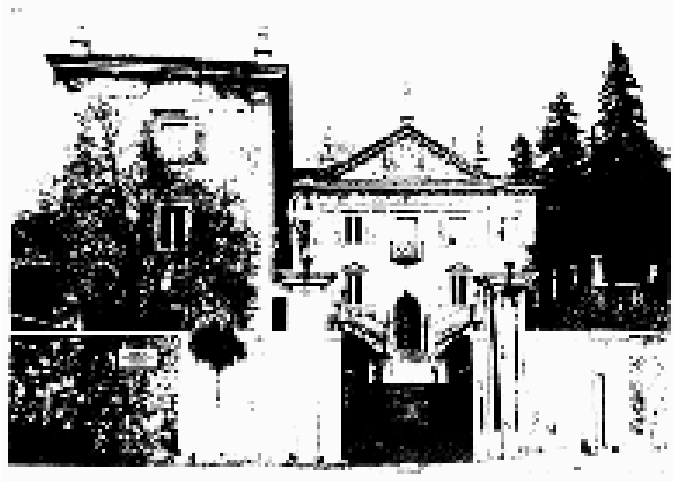,height=7cm,width=10cm}}}
        \end{center}
     \end{figure}
    \HRule
    \begin{center}
        \Large {\bf European Centre for Theoretical Studies in Nuclear
        Physics and Related Areas}
    \end{center}
    \begin{center}
        {\bf Strada delle Tabarelle 286, I--38050 Villazzano (TN),
        Italy}
    \end{center}
    \begin{center}
        {\bf tel.\ +39--461--314730, fax.\ +39--461--935007}
    \end{center}
     \begin{center}
        {\bf e--mail: ectstar@ect.unitn.it, www: http://www.ect.unitn.it}
    \end{center}
    \HRule

\clearpage

\title{Self-energy of $\Lambda$ in finite nuclei}

\author{M.\ Hjorth--Jensen}

\address{ECT*, European Centre for Studies in
Theoretical Nuclear Physics
and Related Areas, Trento, Italy}

\author{A.\ Polls and A.\ Ramos}

\address{Departament d'Estructura i Constituents de la Mat\`eria,
Universitat de Barcelona, Barcelona, Spain}

\author{H.\ M\"{u}ther}

\address{Institut f\"{u}r Theoretische Physik, Universit\"{a}t T\"{u}bingen,
T\"{u}bingen, Germany}

\maketitle

\begin{abstract}
The self--energy of the strange baryon
$\Lambda$ in $^{17}$O is calculated using a microscopic
many--body approach which accounts for correlations beyond the
mean--field or Hartree--Fock approximation. The non-locality and
energy-dependence of the $\Lambda$ self--energy is discussed and the
effects on the bound and scattering states are investigated.
For the nucleon--hyperon interaction, we use the potential models of the
J\"{u}lich and Nijmegen groups.
\end{abstract}


\section{Introduction}

Hypernuclear physics
has received a lot of attention since the early emulsion and bubble
chamber experiments
\cite{juric}
aimed at establishing how the
presence of
a new flavor (strangeness) broadens the knowledge achieved by the
conventional field of nuclear physics and helps in understanding the
breaking of SU(3) symmetry.
Although
major achievements in hypernuclear physics have been taken at a very
slow pace due to limited statistics,
the in--flight $(K^-,\pi^-)$ counter experiments carried out at
CERN \cite{bertini,bruckner} and Brookhaven  \cite{chrien} 
revealed a
considerable amount of hypernuclear
features, such as small spin--orbit strength, increased validity of
single--particle motion of the $\Lambda$, narrow widths
of $\Sigma$--hypernuclei, etc,
injecting
a renewed interest in the field.  
Since then, the experimental facilities have been upgraded and
experiments using the $(\pi^+,K^+)$ and $(K^-_{\rm stopped},\pi^0)$
reactions are being conducted at the Brookhaven AGS and KEK
accelerators with higher beam intensities and improved energy
resolution. Moreover, the photo-- and electro--production of strangeness
will be studied at CEBAF \cite{CEBAF}. It is expected
that the new improved
experimental data will bring the field of hypernuclear physics to a
stage in which major advances can be made.

 From the theoretical side, one of the goals of hypernuclear research
is to relate the hypernuclear observables to the bare hyperon-nucleon
($YN$) interaction. The experimental
difficulties associated to the short lifetime of 
hyperons and low intensity beam fluxes have limited
the number of $\Lambda N$ and $\Sigma N$ scattering
events to less than one thousand \cite{scat1,scat2,scat3,scat4,scat5},
not being enough to fully constrain the $YN$ interaction. At
present, there are two meson-exchange
$YN$ potentials:
that of the Nijmegen group \cite{nijmegen}, where the corresponding
baryon--baryon-meson vertices are subject to strict SU(3) symmetry, and
that
of the J\"ulich group \cite{juelich}, which assumes a stronger SU(6)
symmetry and therefore all the coupling constants at the strange
vertices can be related to the $NN$ coupling constants. Although both
models are able to describe the $YN$ scattering data, their
spin--isospin structure is very different.
Therefore,
more data on $YN$ scattering, especially the measurement of spin
observables, are highly desirable.
In the lack of such data, alternative information can be obtained
from the study of hypernuclei.
One possibility is to
focus on light hypernuclei, such as $^3_{\Lambda}$He,
$^3_{\Lambda}$H and $^4_{\Lambda}$He, which can be treated
``exactly'' solving 3-body Faddeev \cite{gloeckle1,gloeckle2} and
4-body Yakubovsky \cite{gibson} equations. However, the power of
these techniques
is limited by the scarce amount of spectroscopic data.
Only the ground state energies and a
particle--stable excited
state for each A=4 species can be used to put further constraints on
the $YN$ interaction. Another possibility is the study of hypernuclei
with larger masses. They can
be reasonably well described by a shell--model picture and, on the
other hand, provide
a substantial amount of hypernuclear excited states.
Obtaining
spectroscopic data at high resolution is one of the ultimate goals of
the hypernuclear program at CEBAF.
Hypernuclear structure calculations must be performed with
an effective $YN$ interaction ($G$--matrix) obtained from the free
$YN$ potential by
solving a Bethe--Goldstone equation. The comparison with
data will help to elucidate which pieces of
of the bare $YN$ potential
need to be changed for a better agreement with experiment.
Some shell--model calculations have already been performed
using a nuclear matter
$YN$ $G$--matrix \cite{bando,yamamoto} or a $G$--matrix calculated
directly in finite nuclei \cite{kuo,halderson}. It
has been observed that the various potentials predict a quite
different positioning of the excited energy levels, showing the
suitability of this approach as a tool to constrain the $YN$
interaction.

The aim of this work is to study the self-energy of the $\Lambda$ in
$^{17}_{\Lambda}$O using a microscopic many--body approach. Our starting point
is a nuclear matter $G$--matrix at a fixed
energy and density, which is used to calculate the $G$--matrix for the
finite nucleus including corrections up to second order. This second--order
correction, which assumes harmonic oscillator states for the occupied
(hole) states and plane waves for the intermediate unbound particle
states, incorporates the correct energy and density dependence of the
$G$--matrix.
The hypernuclear structure calculations of Ref.  \cite{yamamoto} take
the nuclear matter $G$--matrix at the Fermi momentum $k_F$ (density) that
reproduces the binding energy of the $\Lambda$ in the hypernucleus
under study.
Our second order finite nucleus calculation eliminates the need to
choose such an effective
Fermi momentum for each single-particle state and
hypernucleus. In this sense it is comparable
with the finite nucleus calculations of Refs.  \cite{kuo} and
 \cite{halderson}.

Using this $G$--matrix we determine a self--energy for the $\Lambda$,
which is non--local and depends on the energy of the hyperon. Solving
the Schr\"{o}dinger equation with this self--energy we are able to 
determine the single particle energies and wave functions of the bound
hyperon. Our approach also provides automatically
the real and imaginary part of the hyperon optical potential
at positive energies and, therefore,
allows to study the hyperon--nucleus scattering properties.
Because of the large momentum transfer, the $(\gamma,K^+)$ reaction at
CEBAF \cite{CEBAF2} and $(\pi^+,K^+)$ reaction at KEK \cite{KEK} give
rise to a significant
quasifree production, i.e.\ hyperons in the continuum. The hyperon may
then scatter from nucleons or nuclei composing the target material. It
is then clear that, in order to analyze such data, a knowledge of the
hyperon--nucleus optical potential will be useful and, at present,
very few calculations exist.
In Ref.  \cite{bando2}, an energy dependent local $\Lambda$-nucleus
potential is calculated from a nuclear matter $G$--matrix by applying a
finite range Local Density Approximation. The hyperon--nucleus optical
potential has also been derived from the nucleon--nucleus potential
using Dirac phenomenology and symmetry considerations to relate the
coupling constants in the strange sector to the non-strange
ones \cite{mares}.

In Section 2 we outline our method to calculate the self--energy of
the $\Lambda$ in a finite nucleus and discuss how it is used
in a Schr\"odinger equation to derive the corresponding eigenvalues
and single--particle properties. Our results are presented in 
Section 3 with a special emphasis on the discussion of the non-locality
and energy-dependence of the $\Lambda$ self--energy.
Some concluding remarks are given in Section 4.

\section{Computational details}

\subsection{Evaluation of the $\Lambda$ self-energy}

\begin{figure}
       \setlength{\unitlength}{1mm}
       \begin{picture}(100,180)
       \put(25,0){\epsfxsize=12cm \epsfbox{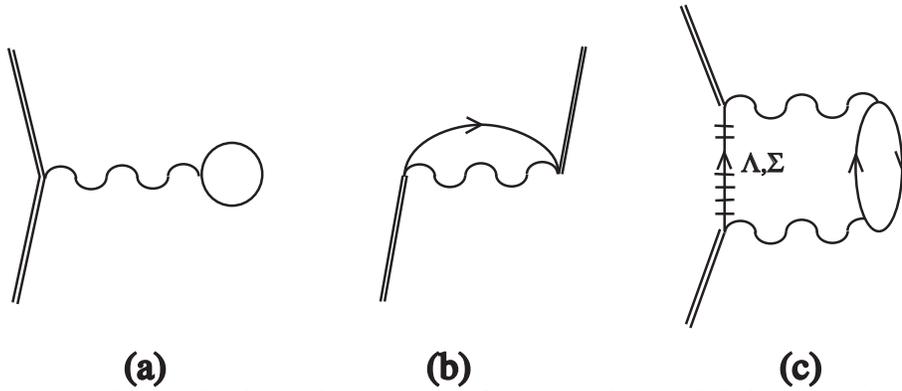}}
       \end{picture}
   \caption{Diagrams through second order in the interaction $G_{YN}$
   (wavy line) included in the evaluation of the self--energy of the
   $\Lambda$. Diagrams
   (a) and (b) represent the direct and the exchange Hartree--Fock terms,
   while (c) is an example of the second order
   two--particle--one--hole diagram. Note that the double external lines
represent a $\Lambda$ while the ``railed'' line in (c)
refers to a intermediate $\Lambda$ and $\Sigma$ hyperon.}
   \label{fig:fig1}
\end{figure}

The self--energy of the $\Lambda$ is evaluated including the
diagrams displayed in Fig.~\ref{fig:fig1}. The wiggly interaction lines
in this figure refer to a $G$--matrix approach for the $YN$ interaction
in nuclear matter calculated at a fixed baryon density and starting
energy. Therefore, as a first step, we calculate this 
$G_{YN}$--matrix in a basis for the two-particle states defined
in terms of relative and center-of-mass momenta 
\begin{eqnarray*}
 {\bf k} &=&\frac{M_N{\bf k_Y}-M_Y{\bf k_N}}{M_N+M_Y}, \\
  {\bf K}&=&{\bf k_N}+{\bf k_Y} \ ,
\end{eqnarray*}
where $M_N$ is the nucleon mass and $M_Y$ the mass of the hyperon
which can be either a $\Lambda$ ($M_\Lambda$) or a $\Sigma$
($M_\Sigma$). The use of an angle--averaged Pauli operator,
see e.g.\ Ref.\  \cite{reuber},
allows us to perform a partial wave decomposition. In terms of the
quantum numbers of the relative and center--of--mass motion (RCM) the
Bethe-Goldstone equation reads
\begin{eqnarray}
\lefteqn{\left\langle k'l'KL({\cal J})S T_z\right |
      G_{YN}\left | k''l''KL({\cal J})S' T_z \right\rangle
      =} \hspace{2cm}\nonumber\\
  && \left\langle k'l'KL({\cal J})S T_z\right |
      V_{YN}\left | k''l''KL({\cal J})S' T_z \right\rangle
      \nonumber \\
      &\quad +& {\displaystyle
      \sum_{l}\sum_{Y=\Lambda\Sigma}\int k^{2}dk}
      \left\langle k'l'KL({\cal J})S T_z\right |
      V_{YN}\left | klKL({\cal J})S' T_z \right\rangle \nonumber \\
      && \times\left\langle klKL({\cal J})S T_z\right |
      G_{YN}\left | k''l''KL({\cal J})S' T_z \right\rangle\nonumber \\
      && \times \frac{Q(k,K)}{\omega_{NM} -\frac{K^2}{2(M_N+M_Y)} -
      \frac{k^2(M_N+M_Y)}{2M_NM_Y}-M_Y+M_\Lambda} \ ,
   \label{eq:gmat}
\end{eqnarray}
where $Q$ is the nuclear matter Pauli operator,
$V_{YN}$ is the $YN$ potential, and
$\omega_{NM}$ is the nuclear matter starting
energy.
The variables $k$, $k'$, $k''$ and $l$, $l'$,
$l''$ denote relative momenta and angular momenta, respectively,
while
$K$ and $L$ are the quantum numbers of the center-of-mass
motion. Further, ${\cal J}$, $S$ and $T_z$ represent the total angular
momentum, spin and isospin projection,
respectively. In our calculations we consider nuclear matter with a Fermi
momentum $k_{F}$ of 1.4 fm$^{-1}$ and a starting energy $\omega_{NM}$ =
--80 MeV, which is a averaged value for the sum of the
single-particle energies of a bound nucleon and a $\Lambda$ at this
density. 

Starting from this $G_{YN}$--matrix in the RCM
system, one can obtain the $G_{YN}$--matrix in the laboratory system
through appropriate transformation coefficients  \cite{bbmp92,hbmp93}.
With these coefficients,
the expression for a two--body wave function in momentum space
using the lab coordinates can be written as
\footnote{Note the distinction between
$k_a$ and $k$ and $l_a$ and $l$. With the notation $k_a$
or $l_a$ we will refer to the quantum numbers of the single--particle
state, whereas $l$ or $k$ without subscripts refer to the coordinates
of the relative motion.}

\begin{equation}
   \begin{array}{ll}
     &\\
     \left | (k_al_aj_at_{z_a})(k_bl_bj_bt_{z_b})JT_z\right \rangle =&
      {\displaystyle \sum_{lL\lambda S{\cal J}}}\int k^{2}dk\int K^{2}dK
      \left\{\begin{array}{ccc}
      l_a&l_b&\lambda\\\frac{1}{2}&\frac{1}{2}&S\\
      j_a&j_b&J\end{array}
      \right\}\\&\\
      &\times (-1)^{\lambda +{\cal J}-L-S}
      \hat{{\cal J}}\hat{\lambda}^{2}
      \hat{j_{a}}\hat{j_{b}}\hat{S}
      \left\{\begin{array}{ccc}L&l&\lambda\\S&J&{\cal J}
      \end{array}\right\}\\&\\
      &\times \left\langle klKL| k_al_ak_bl_b\right\rangle
      \left | klKL({\cal J})SJT_z\right \rangle ,
   \end{array}
\end{equation}
where the term $\left\langle klKL| k_al_ak_bl_b\right\rangle$
is the transformation coefficient from the RCM system to the lab system
 defined in Refs.\  \cite{kkr79,wc72}.

Typical matrix elements needed in the calculation are
\begin{equation}
   \left\langle (k_al_aj_at_{z_a})(n_bl_bj_bt_{z_b})JT_z\right |
    G_{YN}\left | (k_cl_cj_ct_{z_c})(n_dl_dj_dt_{z_d})JT_Z \right\rangle,
\end{equation}
for the Hartree--Fock diagram of Fig.~\ref{fig:fig1}(a), or
\begin{equation}
   \left\langle (k_al_aj_at_{z_a})(n_bl_bj_bt_{z_b})JT_Z \right |
    G_{YN}\left | klKL({\cal J})ST_z \right\rangle ,
\end{equation}
appearing in the second--order diagram of Fig.~\ref{fig:fig1}(b).
The calculation of these matrix elements
require the knowledge of two--body states in a mixed
representation with harmonic oscillator (h.o.) and plane wave states given by
\begin{equation}
\hspace{-0.3cm}\left | (n_al_aj_at_{z_a})(k_bl_bj_bt_{z_b})JT_z \right \rangle=
    \int k_a^2dk_aR_{n_al_a}(\alpha k_a)
    \left | (k_al_aj_at_{z_a})(k_bl_bj_bt_{z_b})JT_z \right \rangle,
\end{equation}
where
$k_al_aj_a$ and $n_al_aj_a$ are shorthands
for plane wave and h.o.\ functions, respectively, and
$\alpha$ is the oscillator parameter which is set to 1.72 fm, a value
which is appropriate to describe the single--particle wave functions of
the bound nucleons in $^{17}_{\Lambda}$O, the system we are going to discuss.
The quantum
numbers $l_{a,b}$,
$j_{a,b}$ and $t_{z_{a,b}}$ are
the single--particle orbital and total angular momenta and isospin
projections, respectively.
The two--body state
is represented by the quantum numbers of the total angular momentum
$J$ and isospin projection $T_z$.

Finally, we are then ready to set up the equations for the
various diagrams evaluated in this work. The direct and exchange
contributions to the HF approximation, displayed in Fig.\ \ref{fig:fig1}(a)
and (b), yield a real and energy-independent the self-energy. It is
given as
\begin{eqnarray}
\lefteqn{  {\cal V}_{HF}(k_{\Lambda}k_{\Lambda}'l_{\Lambda}j_{\Lambda}
   t_{z_{\Lambda}})\quad = \quad \frac{1}{\hat{j_{\Lambda}}^2}
  {\displaystyle \sum_{J}\sum_{n_hl_hj_ht_{z_h}}}\hat{J}^2}\hspace{1.5cm}
\nonumber \\
  & \times & \left\langle
(k_{\Lambda}l_{\Lambda}j_{\Lambda}t_{z_{\Lambda}})
  (n_hl_hj_ht_{z_h})JT_z \right |
   G_{YN}\left | (k_{\Lambda}l_{\Lambda}j_{\Lambda}t_{z_{\Lambda}})
   (n_hl_hj_ht_{z_h})JT_z \right \rangle ,
\label{eq:hf}
\end{eqnarray}
where $\hat{x}=\sqrt{2x+1}$ and $n_hl_hj_ht_{z_h}$ are the quantum numbers
of the nucleon hole states.
The variables
$l_{\Lambda}$, $j_{\Lambda}$, $t_{z_{\Lambda}}$ are the
orbital angular momentum, total angular momentum and isospin
projection ($t_{z_\Lambda} = 0$) of the incoming/outgoing
$\Lambda$, and
$k_{\Lambda}$ ($k_{\Lambda}'$) the incoming (outgoing)
particle momentum.

To calculate the contributions from the two--particle--one--hole ($2p1h$)
diagrams like the example displayed in Fig.\
\ref{fig:fig1} (c)
we evaluate the imaginary part first. The real part is obtained
through the dispersion relation to be defined below.
The analytical expression for the imaginary contribution of the
$2p1h$ diagram, which gives rise to an explicit energy dependence of
the self--energy, is
\begin{eqnarray}
\lefteqn{{\cal W}_{2p1h}(j_{\Lambda}l_{\Lambda}k_{\Lambda}
      k_{\Lambda}'t_{z_{\Lambda}}\omega) \ = 
      {\displaystyle -\frac{1}
      {\hat{j_{\Lambda}}^2}\sum_{n_hl_hj_ht_{z_h}}
      \sum_{J}\sum_{lLS{\cal J}}\sum_{Y=\Lambda\Sigma}\int k^{2}dk
      \int K^{2}dK\hat{J}\hat{T}}}\hspace{1cm}\nonumber \\
      &\times& \left\langle (k_{\Lambda}'l_{\Lambda}j_{\Lambda}
      t_{z_{\Lambda}})(n_hl_hj_ht_{z_h})JT_z\right |
      G_{YN}\left | klKL({\cal J})SJT_z \right \rangle\nonumber \\
      &\times& \left\langle klKL({\cal J})SJT_z \right | G_{YN}
      \left | (k_{\Lambda}l_{\Lambda}j_{\Lambda}
      t_{z_{\Lambda}})(n_hl_hj_ht_{z_h})JT_z \right \rangle \nonumber
      \\ &\times& \pi\delta
      \left(\omega +\varepsilon_h -\frac{K^2}{2(M_N+M_Y)} -
      \frac{k^2(M_N+M_Y)}{2M_NM_Y}-M_Y+M_\Lambda \right) \ ,
   \label{eq:2p1h}
\end{eqnarray}
where $\omega$ is the energy of the $\Lambda$ measured with respect
to the $\Lambda$ rest mass.
The single--hole energies $\varepsilon_{h}$ are set equal to the
experimental single-particle energies in $^{16}$O.
The quantities $klKL({\cal J})SJT_z$ are the
quantum numbers of the intermediate $YN$ state.
In the above sum over intermediate nucleon and hyperon states, we have to
account for the fact that the nucleon particle states should be orthogonal
to the hole states. This is done following the
local density approximation for the orthogonalization of the intermediate
nucleon particle states of Ref.\  \cite{bbmp92}.

The  contributions to the real part of the self-energy from Eq.\
(\ref{eq:2p1h}) can be obtained through the following
dispersion relation
\begin{equation}
   {\cal V}_{2p1h}(j_{\Lambda}l_{\Lambda}k_{\Lambda}k_{\Lambda}'
   t_{z_{\Lambda}} \omega)=
   \frac{P}{\pi} \int_{-\infty}^{\infty}
   \frac{{\cal W}_{2p1h}(j_{\Lambda}l_{\Lambda}k_{\Lambda}
    k_{\Lambda}'
    t_{z_{\Lambda}} \omega')}{\omega'-\omega} d\omega',
    \label{eq:disprel}
\end{equation}
where $P$ means a principal value integral. Since  ${\cal W}_{2p1h}$
is different from zero only for positive values of
$\omega'$ and its diagonal matrix elements are negative,
this dispersion relation  implies that the diagonal elements of
${\cal V}_{2p1h}$ will be attractive for
negative values of $\omega$. This attraction should increase for small
positive energies. It will eventually decrease and become repulsive
only
for large positive values of the energy of the interacting $\Lambda$.

The reader should observe that the Hartree--Fock contribution of
Eq.\ (\ref{eq:hf}) has been obtained with a $YN$ $G$--matrix
calculated in nuclear matter. Therefore, the Hartree--Fock contribution
contains
already $2p1h$ terms like those displayed in Fig.\ \ref{fig:fig1}(c),
but calculated for nuclear matter. 
To avoid this
double-counting of intermediate $YN$ states, we subtract
from the
real part of the $\Lambda$ self-energy
a correction term
\begin{eqnarray}
\lefteqn{\hspace{-0.5cm}{\cal V}_{c}(j_{\Lambda}l_{\Lambda}k_{\Lambda}
      k_{\Lambda}'t_{z_{\Lambda}}) \ = \
      {\displaystyle \frac{1}
      {\hat{j_{\Lambda}}^2}\sum_{n_hl_hj_ht_{z_h}}
      \sum_{J}\sum_{lLS{\cal J}}\sum_{Y=\Lambda\Sigma}\int k^{2}dk
      \int K^{2}dK\hat{J}\hat{T}}}\hspace{0.5cm}\nonumber \\
      &\times& \left\langle (k_{\Lambda}'l_{\Lambda}j_{\Lambda}
      t_{z_{\Lambda}})(n_hl_hj_ht_{z_h})JT_z\right |
      G_{YN}\left | klKL({\cal J})SJT_z \right \rangle\nonumber \\
      &\times& \left\langle klKL({\cal J})SJT_z \right | G_{YN}
      \left | (k_{\Lambda}l_{\Lambda}j_{\Lambda}
      t_{z_{\Lambda}})(n_hl_hj_ht_{z_h})JT_z \right \rangle \nonumber
      \\
      &\times& Q(k,K)\left(\omega_{NM} -\frac{K^2}{2(M_N+M_Y)} -
      \frac{k^2(M_N+M_Y)}{2M_NM_Y}-M_Y+M_\Lambda\right)^{-1} \ ,
   \label{eq:vc}
\end{eqnarray}
where $Q$ is the nuclear matter Pauli operator
and $\omega_{NM}$ is the nuclear matter starting
energy.

In summary, the self--energy of the $\Lambda$ hyperon reads
\begin{equation}
    \Sigma(j_{\Lambda}l_{\Lambda}k_{\Lambda}k_{\Lambda}'\omega)=
    V(j_{\Lambda}l_{\Lambda}k_{\Lambda}k_{\Lambda}'\omega)+
    iW(j_{\Lambda}l_{\Lambda}k_{\Lambda}k_{\Lambda}'\omega),
    \label{eq:self_ener}
\end{equation}
with the real part given by
\begin{equation}
    V(j_{\Lambda}l_{\Lambda}k_{\Lambda}k_{\Lambda}'\omega)=
       {\cal V}_{HF}(j_{\Lambda}l_{\Lambda}k_{\Lambda}k_{\Lambda}')+
       {\cal V}_{2p1h}(j_{\Lambda}l_{\Lambda}k_{\Lambda}k_{\Lambda}'
        \omega)-
       {\cal V}_c(j_{\Lambda}l_{\Lambda}k_{\Lambda}k_{\Lambda}'
       )
       \label{eq:realV}
\end{equation}
and the imaginary part by

\begin{equation}
    W(j_{\Lambda}l_{\Lambda}k_{\Lambda}k_{\Lambda}'\omega)=
       {\cal W}_{2p1h}(j_{\Lambda}l_{\Lambda}k_{\Lambda}k_{\Lambda}'\omega).
       \label{eq:imV}
\end{equation}

\subsection{Solution of the Schr\"odinger equation}

The self--energy of Eq.\ (\ref{eq:self_ener})
can be
used as a single--particle potential  in a
Schr\"o\-din\-ger equation in order to investigate
bound and scattering states of a  $\Lambda$
in a finite nucleus.
The different approximations to the
self--energy, i.e.\ whether we include the $2p1h$
contribution
or not, results in different single--particle hamiltonians.
The Schr\"o\-din\-ger equation is solved by diagonalizing the corresponding
single--particle hamiltonian in a complete basis within a spherical box
of radius $R_{box}$. The radius of the box should be larger than the radius
of the nucleus considered. The calculated observables are independent of the
choice of $R_{box}$, if it is chosen to be around $15$ fm or larger. 
The method is especially suitable for non--local 
potentials defined either in coordinate or in momentum space.

A complete and orthonormal set of regular basis functions within this box is
given by
\begin{equation}
    \Phi_{iljm}({\bf r})= \left\langle {\bf r} | k_i ljm \right\rangle= 
    N_{il} j_l(k_i r) \psi_{ljm}(\theta \phi)
    \label{eq:basis}
\end{equation}
In this equation 
$\psi_{ljm}$ represent the spherical harmonics 
including the spin degrees of freedom and $j_l$ denote 
the spherical Bessel functions for the discrete momenta $k_i$ which fulfill
\begin{equation}
    j_l(k_i R_{box}) =0.
\end{equation}
For the specific case of $l=0$, which we will consider
in this paper, the normalization constant is 
\begin{equation}
N_{i0} = \frac {i \pi 2^{1/2}}{R_{box}^{3/2}} \ .
\end{equation}
In this way, the basis functions defined in Eq.\ (\ref{eq:basis}) are
orthogonal and normalized within the box.  
The single particle hamiltonian for the $\Lambda$,
consisting out of the kinetic energy and the real part of the self--energy
can be evaluated in this basis and the resulting eigenvalue problem
\begin{equation}
     \sum_{n=1}^{N_{max}} \left\langle k_i \right | \frac {k_i^2}
   {2 m_{\Lambda}} 
       \delta_{in} +V(\omega =E_{\Upsilon}) \left | k_n \right\rangle 
 \left\langle k_n | \Upsilon\right\rangle = E_{\Upsilon} \left\langle k_i |
    \Upsilon \right\rangle,
\label{eq:hamil}
\end{equation}
restricted typically to $20$ or $30$ states, can be easily solved. Notice
that a self--consistent process is performed for each eigenvalue, i.e. 
the self--energy needs to be evaluated at the energy of the resulting 
eigenvalue. 
As a first result, one obtains the negative energies for the bound states and 
the corresponding wave functions, which are
expressed in terms of expansion coefficients 
for the basis defined in
Eq.\ (\ref{eq:basis}). Furthermore, one also obtains discrete positive
energies that correspond to scattering states with radial wave functions
which are zero at $r=R_{box}$. Taking into account this fact it is
straightforward to evaluate the phase shifts for those energies. Phase
shifts for different energies can be obtained by varying $R_{box}$, see
e.g.\ the discussion in Ref.\   \cite{bbmp92}.

\section{Results and discussion}

The results presented in this section have been obtained with
two potential models for the free $YN$--interaction $V_{YN}$, namely the
Nijmegen soft--core potential described
in Ref.\  \cite{nijmegen} and the
energy independent version with parameter set B of the
J\"ulich group
 \cite{juelich,reuber}. 
Both potential models are based on meson--exchange theory and include
the relevant low--energy mesons.
However,
contrary to what is the case for the nucleon--nucleon potential,
not all parameters which enter the definition of the various potential
models, like coupling constants, are left as free
parameters to be constrained by the data.
The Nijmegen model imposes SU(3) symmetry on the coupling constants
leaving the pseudoscalar $F/D$ ratio $\alpha$ as a free parameter
adjusted to fit the data ($\alpha=0.355$). Instead, the J\"ulich
model fixes its value to $\alpha=2/5$ by imposing the stronger SU(6)
symmetry. Thus, various
potential models may give different results for various observables.

As an example, we show in Table \ref{tab:tab1} the binding energy
of the $\Lambda$ at rest in nuclear matter at
saturation density ($k_F=1.4$ fm$^{-1}$). 
These results were obtained using a starting energy
$\omega_{NM}=-80$ MeV,
which is approximately the sum of the
single--particle energies of an average nucleon and a $\Lambda$ in
nuclear matter,
and with the center--of--mass momentum of the $YN$ pair
at rest. We employ kinetic energies for the single--particle spectrum
above the Fermi momentum.  The numbers displayed in Table \ref{tab:tab1}
compare well with the nuclear matter calculations
of Reuber {\em et al.}  \cite{reuber} and
Yamamoto {\em et al.\ }  \cite{yamamoto}.
One can see that the Nijmegen potential predicts a binding energy for
the $\Lambda$ (--24.35 MeV), which is about 7 MeV weaker than the
prediction of the J\"ulich model (--31.48 MeV). Even larger differences
are observed if one inspects the contributions of the various partial
waves. While for the case of the J\"ulich potential the predominant
contribution to the binding energy results from the $^3S_{1}$--$^3D_{1}$
coupled channel, the largest term for the Nijmegen potential is
obtained in the $^1S_{0}$ channel. In both interaction models a large
part of the $\Lambda$ binding energy is due to the coupling of the
$\Lambda N$ to the $\Sigma N$ channel. Neglecting this coupling
(numbers listed in parenthesis) would result in unbound $\Lambda$ in
the case of the Nijmegen potential.

The main purpose of this work, however, is to study the $\Lambda$
self--energy in $^{17}_{\Lambda}$O
as well as
the single--particle properties that can be derived from it.
Here we restrict our attention to
the states with orbital angular momentum
$l_{\Lambda}=0$. Unless the comparison turns to be of interest, 
we will present results for the J\"ulich interaction only, since
it yields the correct $\Lambda$ binding energy in nuclear matter as
obtained when extrapolating from $A\to \infty$ to finite hypernuclei
data \cite{bando3}.

\begin{figure}
       \setlength{\unitlength}{1mm}
       \begin{picture}(100,180)
      \put(25,0){\epsfxsize=12cm \epsfbox{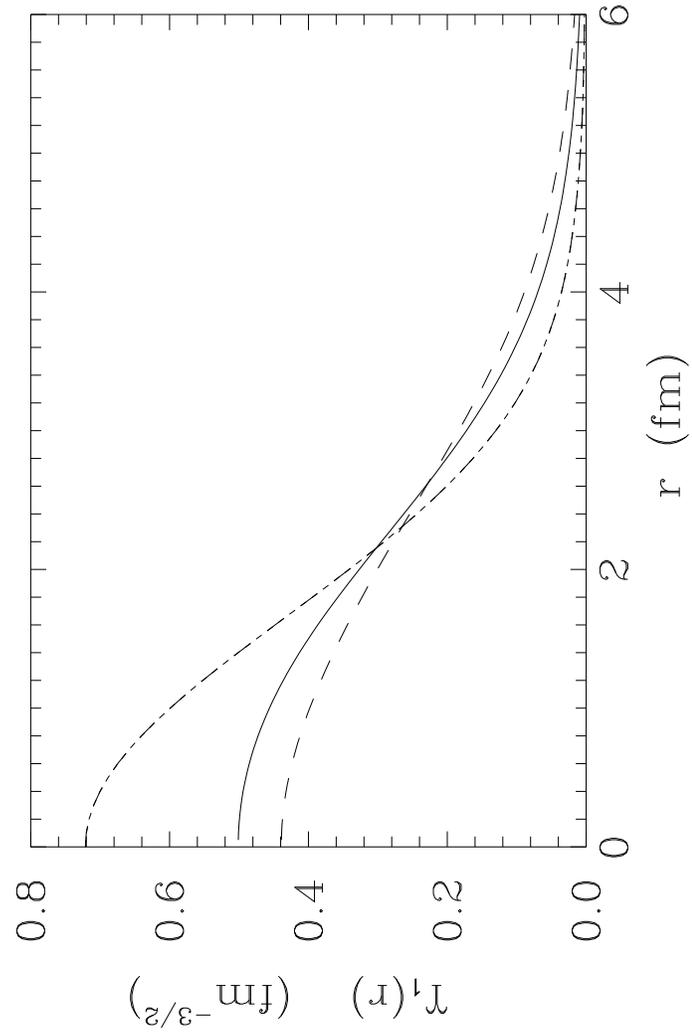}}
       \end{picture}
   \caption{Wave function in $r$--space for the $\Lambda$ in the
            $0s_{1/2}$ state in
            $^{17}_{\Lambda}$O for the J\"ulich (solid line)
            and the Nijmegen (dashed line)
            potentials. For comparison we include the
            single nucleon wave function
            (dash--dotted line) in
           $^{16}$O from Ref.\ [26].}
   \label{fig:fig2}
\end{figure}

We first investigate the properties of the known bound state
of an s--wave  $\Lambda$ in $^{17}_{\Lambda}$O.
In Table \ref{tab:tab2} we display
the binding energy, kinetic energy and root--mean--square radius of the
$\Lambda$ obtained by solving the Schr\"odinger equation for two
different choices of the self--energy. As can be seen from this table,
both potentials give a bound state already
at the Hartree--Fock level. Recall, however, that the label
Hartree--Fock approximation here refers to a calculation in terms of
the nuclear matter $G$--matrix. Similar to the case of nuclear matter
the Nijmegen potential yields smaller binding energies than the
J\"ulich model also for the finite system. This smaller binding energy
predicted by the Nijmegen potential also results in a larger radius for
the bound $\Lambda$ state and a smaller binding energy.
This feature can also
be seen in Fig.\ \ref{fig:fig2}, where we plot the corresponding
$\Lambda$ wave functions. For the sake of comparison we also
plot the wave function of a $0s_{1/2}$ nucleon in
$^{16}$O taken from Ref.\  \cite{hbmp93}.

The second--order correction introduces additional attraction, which
can be understood from the following argument:
The total $2p1h$ contribution is given by ${\cal V}_{2p1h}$, see
Eq.\ (\ref{eq:disprel}),
minus the nuclear matter correction term ${\cal V}_c$
of Eq.\ (\ref{eq:vc}).
Thus, Eq.\ (\ref{eq:realV}) 
introduces the finite nucleus
Pauli operator which is less restrictive than the nuclear matter
Pauli operator at $k_F=1.4$ fm$^{-1}$.
Therefore, by allowing for a larger phase
space in the sum over intermediate
states, the finite nucleus second--order contribution
is more attractive than the corresponding nuclear matter result,
producing an overall attractive second--order correction.
It is important to note that 
the final result ($HF+2p1h$) is  stable with respect to the use of
different starting energies
in the $G$--matrix. In other words, had we used a different value
for the starting energy, $\omega_{NM}$, the $HF$ contribution to
the $\Lambda$ single--particle energy would have had a different
value. However,
the $2p1h$ contribution would have also been different producing
approximately the same final results as shown in Table \ref{tab:tab2}.

One of the advantages
of the present approach is that it also allows the study of
scattering states. Since this is
traditionally
done in terms of energy--dependent local potentials
we now wish to investigate
how well our non--local self--energy obtained
from Eq.\ (\ref{eq:self_ener}) can
be represented by a local potential.
We start by analyzing the Hartree--Fock
contribution, which does not contain any explicit dependence on the energy
of the incoming $\Lambda$. In this approximation the non--locality
arises to some extent from the non--local $G$ interaction but mainly
from the Fock exchange term displayed in Fig.\ \ref{fig:fig1}(b).

The localization of the self--energy is more easily discussed in $r$--space
and therefore it is useful to consider the Fourier--Bessel transform
from the momentum space to $r$--space
\begin{equation}
  {\cal V}_{HF}(l=0,r,r')= \frac {2}{\pi} \int k^2 dk \int {k'}^2
                       dk' j_0(kr) j_0(k' r') {\cal V}_{HF}(l=0,k,k') \ .
  \label{eq:tranbe}
\end{equation}
We  generate a local representation 
of ${\cal V}_{HF}(l=0,r,r')$ by performing an average of
the non--local potential over the coordinate $r'$  weighted with the
radial function of the lowest bound state, $\Upsilon_1(r)$
\begin{equation}
     {\cal V}_{HF}^{loc}(l=0,r)= \frac {\int dr' r'^2 {\cal V}_{HF}(l=0,
                                 r,r') \Upsilon_{1}(r')}{\Upsilon_1(r)}.
     \label{eq:local}
\end{equation}
This procedure ensures that the local potential ${\cal V}_{HF}^{loc}(r)$ will
give rise to the same bound state 
$\Upsilon_1(r)$. Notice that the absence of nodes in the lowest bound
state guarantees the numerical stability of this prescription.

\begin{figure}
       \setlength{\unitlength}{1mm}
       \begin{picture}(100,180)
      \put(25,0){\epsfxsize=12cm \epsfbox{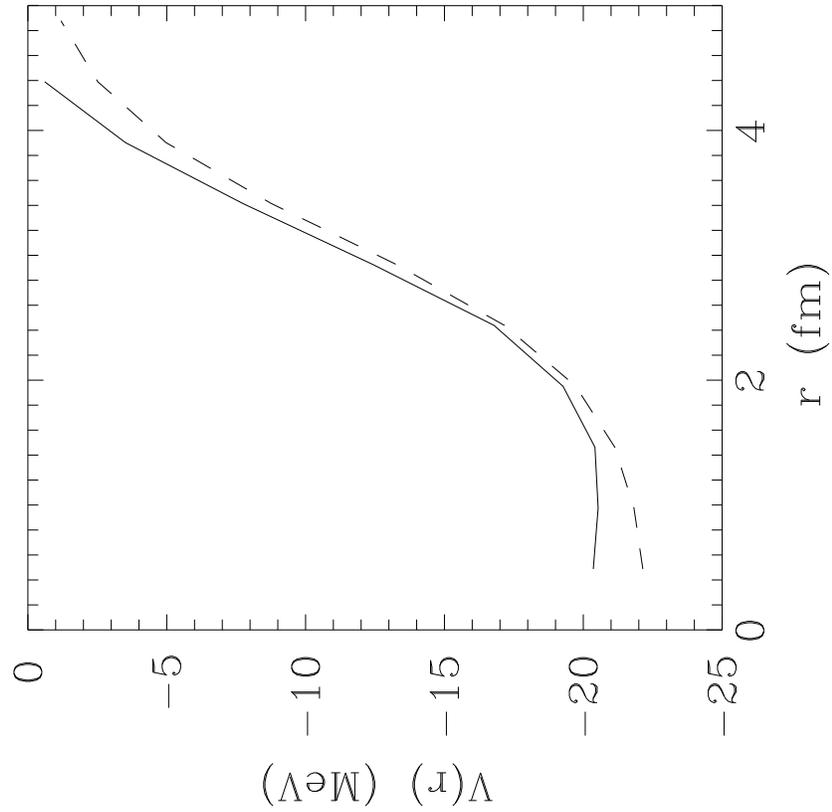}}
       \end{picture}
   \caption{ Local single--particle potentials for a $\Lambda$
             in the $0s_{1/2}$ state in
             $^{17}_{\Lambda}$O employing the J\"ulich potential.
             Solid line represents
              the results obtained from Eq.\ (19) while the
             dashed line is the result
             obtained with the Woods--Saxon parametrization discussed
             in the text.}
   \label{fig:fig3}
\end{figure}

The results of this localization are represented by the solid line
in Fig.\  \ref{fig:fig3}. One can see that this
local representation might, in first approximation, be characterized
by the shape of a Woods--Saxon potential
\begin{equation}
      V_{WS}(r)=\frac{V_0}{1+\exp{(r-R)/a}},
      \label{eq:ws}
\end{equation}
with parameters $V_0=-22.40$ MeV, $R=3.15$ fm and
$a=0.6$ fm, shown by the dashed
line of Fig.\ \ref{fig:fig3}.
These parameters reproduce the binding energy of the
$\Lambda$ and yield an
overlap with the wave function obtained with the Hartree--Fock
self--energy equal to $0.99994$.

\begin{figure}
       \setlength{\unitlength}{1mm}
       \begin{picture}(100,180)
      \put(25,0){\epsfxsize=12cm \epsfbox{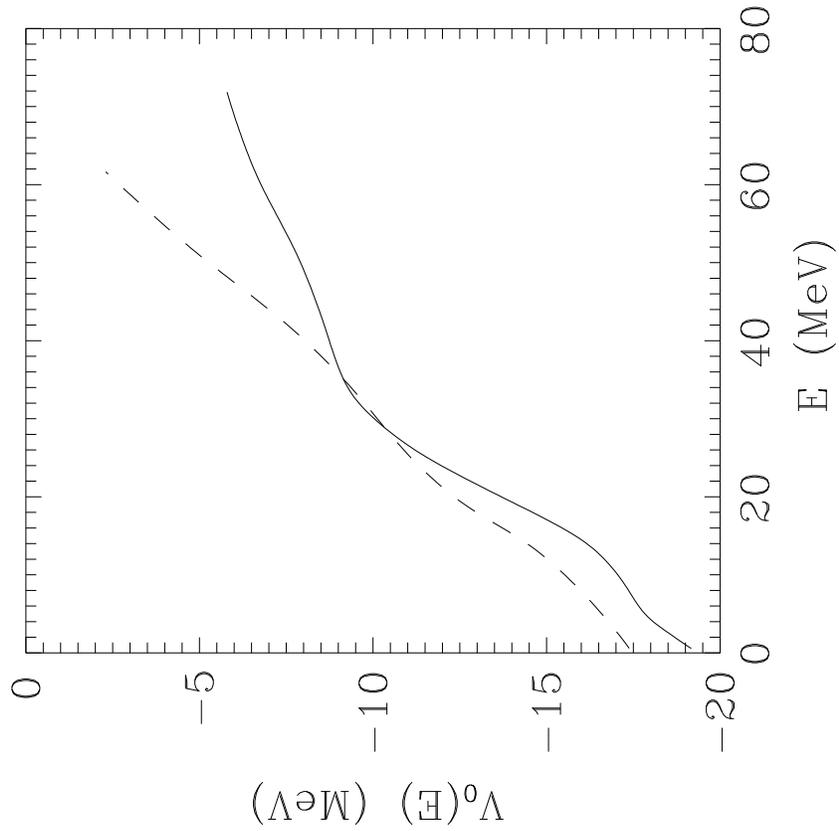}}
       \end{picture}
   \caption{Energy dependence of the depth of the $\Lambda$--nucleus
            Woods--Saxon potential for  a $\Lambda$ in the
           $l=0$ state employing the J\"ulich potential. 
           The dashed line shows 
           the depths resulting from fitting the phase--shifts to those 
           obtained by including only the Hartree--Fock
           diagram to the self--energy.
           Solid line is obtained by including also
           the two--particle--one--hole diagram
           in the evaluation of the self--energy.}
   \label{fig:fig4}
\end{figure}

Having established the radial shape of the local effective
$\Lambda$--nucleus potential for the bound state, we now proceed to
study the region of positive energies. By keeping
$R$ and
$a$ fixed, we allow for an energy dependent depth, $V_0(E)$, chosen to
reproduce the same phase shift at each energy as obtained with the
non--local Hartree--Fock
self--energy. The depth of the Woods--Saxon potential
is shown as a function of energy by
the dashed line in Fig.\ \ref{fig:fig4}.
The behavior of the $\Lambda$--nucleus potential as a function of
energy or as a function of the asymptotic momentum $k$ related to this
energy, can be characterized by an effective $k$--mass $m_{k,\Lambda}$
 \cite{ms91}
\begin{equation}
     \frac{m_{k,\Lambda}(E,r)}{m_{\Lambda}}
=\left(1+\frac{m_{\Lambda}}{k}\frac{dV}{dk}\right)^{-1} =
\left(1+\frac{dV(E,r)}{dE}\right)^{-1}. \label{eq:kmas}
\end{equation}
If we approximate the dashed curve
in Fig.\  \ref{fig:fig3} by a straight line, we obtain a constant
value of this effective mass of about $m_{k,\Lambda}/m_{\Lambda}=0.8$.
Note that this
value is a measure of the momentum dependence or non--locality of the
Hartree--Fock 
self--energy, which is the only source of the effective energy
dependence for the local equivalent potential.

\begin{figure}
       \setlength{\unitlength}{1mm}
       \begin{picture}(100,180)
      \put(25,0){\epsfxsize=12cm \epsfbox{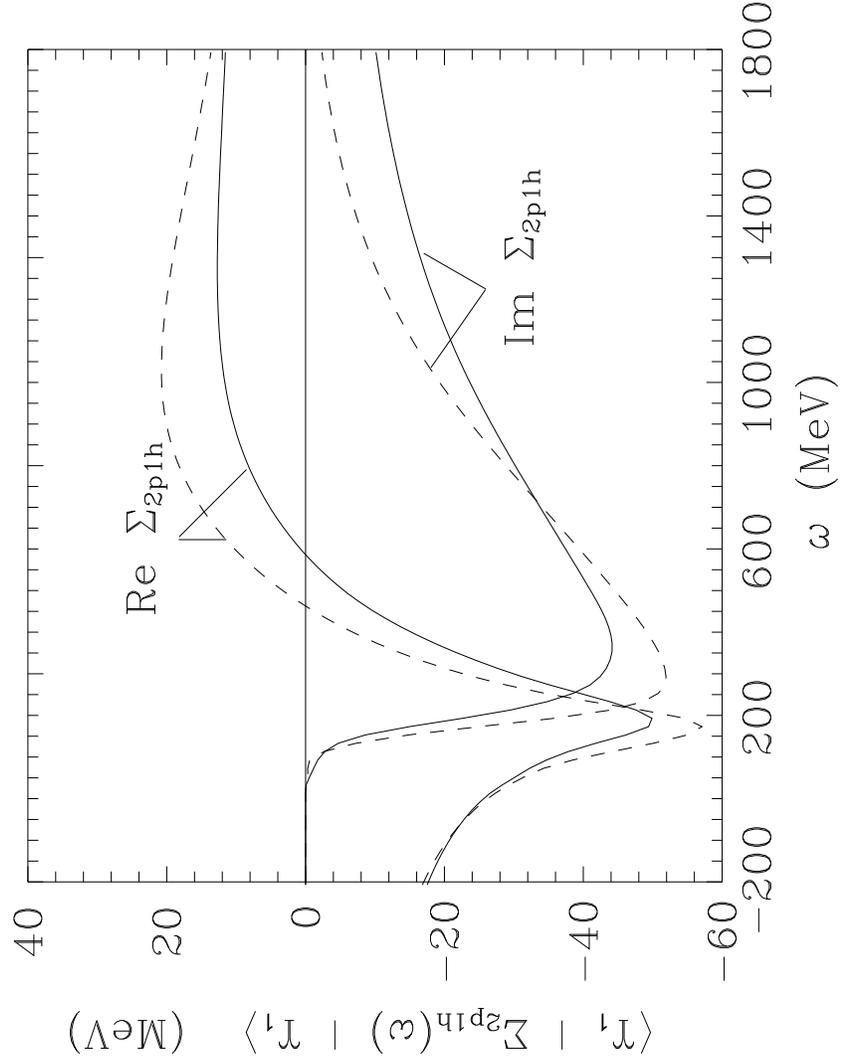}}
       \end{picture}
   \caption{Ground state expectation value of the real and imaginary parts
            of the dispersive term of the $\Lambda$ self--energy as functions
            of $\omega$.
            Solid lines are results obtained with the J\"ulich
            potential while dashed lines are those of the Nijmegen potential.}
   \label{fig:fig5}
\end{figure}

Beside this effective energy dependence, the second order contribution
to the $\Lambda$ self--energy, ${\cal V}_{2p1h}$, also yields an explicit
energy dependence,
which is tied to the coupling to intermediate $2p1h$ states
as shown in Eq.\ (\ref{eq:2p1h}).
The energy dependence of this dispersive correction can be
better visualized by calculating
the expectation value of the second order contribution to the
self--energy in the $l_{\Lambda}=0$
ground state,
$\langle \Upsilon_1 \mid {\cal V}_{2p1h} (\omega) + i
{\cal W}_{2p1h}(\omega) \mid
\Upsilon_1 \rangle $.
In Fig.\ \ref{fig:fig5}
the real and imaginary parts
of the dispersive term of the self--energy 
are shown as functions of energy.
The solid line
shows the results obtained with J\"ulich potential while the dashed
line corresponds to the Nijmegen potential.
As can be seen from this
figure, the absolute value of the imaginary component ${\cal W}_{2p1h}$
is slightly larger for the Nijmegen potential as compared to the
J\"ulich model at small energies $\omega$. Therefore the dispersion
relation of Eq.\ (\ref{eq:disprel}) leads to a stronger energy dependence
in the real part of the self--energy around $\omega = 0$ in the case of
the Nijmegen potential. It is this energy dependence, which leads to
the larger $2p1h$ correction for the bound state in the case of the
Nijmegen potential (see Table \ref{tab:tab2}).

The non--locality of the real part of the complete self--energy (see
Eq.\ (\ref{eq:realV}))
can also be characterized by a local Woods--Saxon with an energy
dependent depth, as
done with the Hartree--Fock term. The energy dependence of this
depth
is represented by the solid line of Fig.\ \ref{fig:fig4}.
The difference between the solid and dashed lines is a measure of
the non--locality and the explicit energy dependence of the dispersive
correction. In the range of energies shown in Fig.\ \ref{fig:fig4}, the
explicit energy dependence should provide attraction increasing with
energy, as can be inferred from the real part of the self--energy plotted
in Fig.\ \ref{fig:fig5}. This would imply that the depth $V_{0}$ of the
local equivalent potential representing the whole self--energy should be
more attractive than the one representing the Hartree--Fock
approximation. However, due to the non--locality of the second order
terms (${\cal V}_{2p1h}-{\cal V}_c$), this attraction can clearly be
seen in Fig.\ \ref{fig:fig4} only at higher energies.
The effective mass characterizing the explicit energy dependence as
well as the momentum dependence of the full self--energy can be
represented as a product of a $k$--mass as defined in Eq.\
(\ref{eq:kmas}) and the energy-dependent mass \cite{ms91}.
This total effective mass can be calculated approximately from the
depth $V_{0}$ of the local equivalent potential by
\begin{equation}
     \frac{m^*_{\Lambda}(E)}{m_{\Lambda}} =
\left(1-\frac{dV_{0}(E)}{dE}\right).
\end{equation}
This effective mass is no longer constant in the range of energies
considered in Fig.\ \ref{fig:fig4}. The value of $m^*/m$ ranges from
about 0.7 at low energies to 0.9 at energies above 50 MeV.

\begin{figure}
       \setlength{\unitlength}{1mm}
       \begin{picture}(100,180)
      \put(25,0){\epsfxsize=12cm \epsfbox{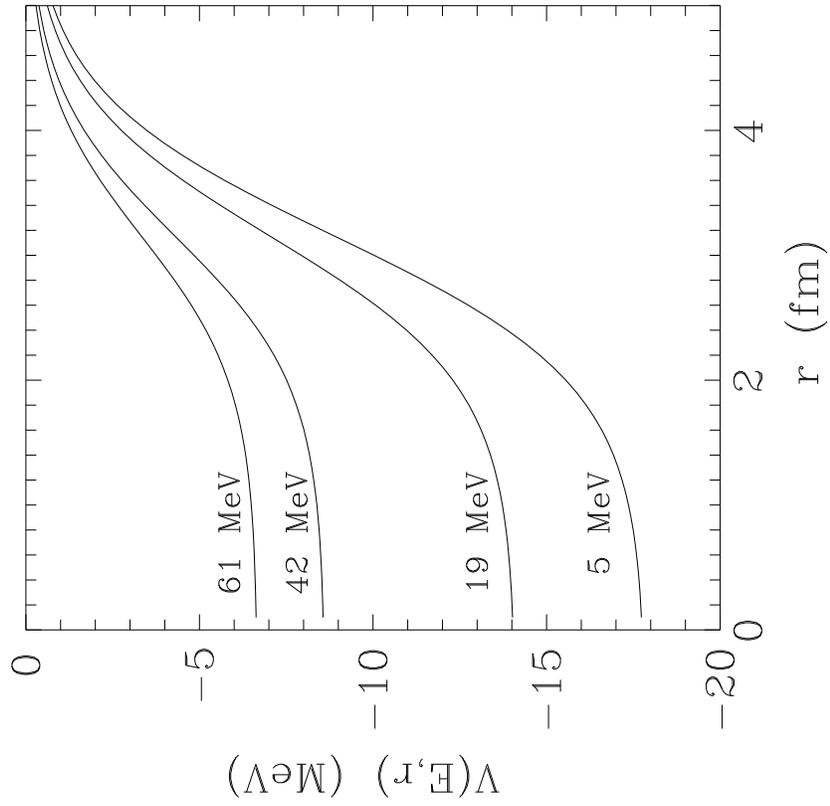}}
       \end{picture}
   \caption{Local $\Lambda$--nucleus Woods--Saxon potential
          at different $\Lambda$
          energies for a $\Lambda$ in the $l=0$ state.}
          \label{fig:fig6}
\end{figure}

The real part of the
$\Lambda$-nucleus optical potential for $l_\Lambda=0$ is
shown as a function of $r$ in Fig.\ \ref{fig:fig6} for several energies
of the incoming $\Lambda$.
Our optical potential exhibits substantial differences with that
of Ref.  \cite{bando2} obtained from a nuclear matter G--matrix and the
local density approximation employing the Nijmegen model D
interaction. The Nijmegen model D gives a nuclear matter $\Lambda$
binding energy of $-40.5$ MeV, the attraction being concentrated
mainly in the $^3S_1-^3D_1$ channel \cite{yamamoto}. It is therefore
more similar
to the potentials of the J\"ulich group than to the Nijmegen soft-core
model.
Figure \ref{fig:fig6} shows that in the energy range $5-60$ MeV the depth
of our optical potential changes by about 10 MeV, whereas in a
similar energy range the depth of the potential obtained in
Ref.\  \cite{bando2} (see their Fig. 4) varies by about half as much.
Moreover, our potential is much shallower. At
$\omega=40$ MeV, for instance, we find a depth of about $-9$ MeV while
a value of $-25$ MeV is obtained in Ref.\  \cite{bando2}. Only part of
this attraction (about 5 MeV) can be attributed to the more
attractive nuclear
matter $G$--matrix obtained with the Nijmegen D model. Part of the
difference can also be attributed to the wider shape of our
equivalent Woods--Saxon. We have tried narrower Woods--Saxon potentials
that still
reproduce reasonably well the binding energy and wave function of the
bound state, but the depth at 40 MeV decreased at most by 3
MeV.
The remaining difference must be attributed to the different methods
used and, therefore, we can conclude that the finite nucleus
calculation presented here gives a substantially less attractive
optical potential than the local density calculation of Ref.\
 \cite{bando2}.

\begin{figure}
       \setlength{\unitlength}{1mm}
       \begin{picture}(100,180)
      \put(25,0){\epsfxsize=12cm \epsfbox{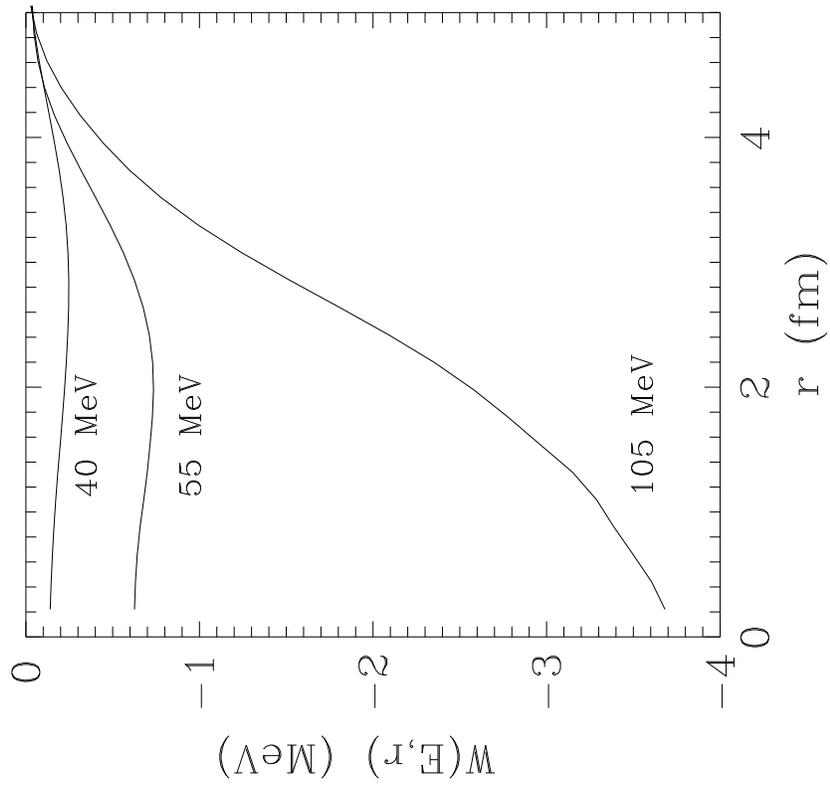}}
       \end{picture}
   \caption{Imaginary part of the optical potential at different $\Lambda$
          energies for a $\Lambda$ in the $l=0$ state.}
          \label{fig:fig7}
\end{figure}

The imaginary--part of the self-energy ${\cal W}_{2p1h}$ as calculated
from Eq.\ (\ref{eq:2p1h}) is also non--local. To obtain a local
representation we follow the same procedure as for the real part
shown in Eq.\ (\ref{eq:local}).
The resulting imaginary part of the $\Lambda$--$^{16}$O optical
potential
is shown in Fig.\ \ref{fig:fig7} for various values of the incoming
$\Lambda$ energy. The imaginary part becomes deeper with increasing
energies and it changes from a slightly surface--peaked shape at
40 MeV (although not very well visible in the scale of Fig.\
\ref{fig:fig7}) to a center--peaked shape at 105 MeV. These features
were also observed in the results of Ref.\  \cite{bando2}
although, again,
we obtain a much shallower depth for the imaginary part of
the $\Lambda$ optical potential.
At 40 MeV, for instance, we obtain a depth of about $-0.2$ MeV while
the value obtained in Ref.\  \cite{bando2} is $-2.4$ MeV.

 From this analysis we conclude that, in order to derive hypernuclear
properties, microscopic calculations of
the effective interaction must be combined with a
reliable treatment of the finite hypernucleus under consideration
since
the single--particle $\Lambda$ properties (single--particle energy,
wave functions, optical potential) are sensitive to the particular
model. This is especially relevant at present for the analysis of
the high resolution
hypernuclear spectroscopy experiments which are going to be conducted
at CEBAF.

\section{Conclusions}

In this work we have calculated the self--energy of a
$\Lambda$ in the  finite nucleus $^{17}_{\Lambda}$O using
a microscopic many--body approach which allows to study
the energy dependence as well as the momentum dependence
of the self--energy for a finite nucleus, 
starting with realistic modern potential
models for the hyperon--nucleon (YN) interaction. As models 
for the YN--interaction we have used the meson--exchange 
models of the Nijmegen  \cite{nijmegen} and J\"ulich groups
 \cite{juelich}. The J\"ulich potential yields a more attractive
$\Lambda$ single--particle energy in nuclear matter than
the Nijmegen potential. The same qualitative pattern is repeated
in the finite nucleus calculation, where the single--particle 
energy for  a $\Lambda$ in the 
$0s_{1/2}$ state in  $^{17}_{\Lambda}$O is $-11.83$ MeV and 
$-7.38$ MeV with J\"ulich and Nijmegen potentials, respectively.
The experimental estimate from Ref.\  \cite{bando3} is
$-12.5$ MeV. 
The self--energy is in turn used to derive an optical potential
for the $\Lambda$--nucleus system. It turns out that the real and
imaginary parts of our 
non-local optical potential can easily be approximated by the 
shape of a Woods--Saxon potential with a depth depending on energy.
This local equivalent potential is less attractive than 
those derived from the local--density calculation of 
Ref.\  \cite{bando2}.  

\section{Acknowledgments}
We thank A.\ Reuber and V.\ Stoks for their help regarding the
use of the $YN$ potentials.
This work has been supported by the Research Council of Norway,
the Istituto Trentino di Cultura, Italy, 
the Spanish research council through DGICYT-grant No.\ PB92-0761
and the EU through contract No. CHRX-CT-93-0323. One of us (H.M.)
would like to thank the University of Barcelona for the hospitality
and the generous support of his visit.


\clearpage


\begin{table}
\begin{center}
\caption{Partial wave contributions to the
binding energy of the $\Lambda$ in nuclear matter for the J\"{u}lich and
Nijmegen potentials at Fermi momentum $k_F=1.4$ fm$^{-1}$.
Numbers in parentheses refer to the case
when the coupling to intermediate $\Sigma N$ is omitted in the calculation of
the $\Lambda N$ $G$--matrix. The total numbers include partial waves
with total angular momentum $J\leq 4$.
All entries in MeV.}
\begin{tabular}{ccccccc}\hline
& $^1S_0$ & $^3S_1$--$^3D_1$ & $^3P_0$ & $^1P_1$--$^3P_1$ & $^3P_2$--$^3F_2$&
Total\\\hline
J\"ulich & -0.6 & -33.95&0.59&3.04&0.093&-31.48 \\
         & (1.61) & (-19.98) & (0.62) & (3.31) & (0.25) & (-11.93) \\

Nijmegen & -14.99 & -8.17 & 0.37 & 3.54 & -3.88 & -24.35 \\
         & (-13.84) & (13.63) & (0.43) & (4.36) & (-2.92) & (0.57) \\
         \hline
\end{tabular}
\end{center}
\label{tab:tab1}
\end{table}

\begin{table}
\begin{center}
\caption{Single--particle energy ($\varepsilon_{\Lambda}$),
mean--square radius ($rms$) and kinetic energy ($T$)
for a $\Lambda$ in the $0s_{1/2}$ state
of $^{17}_{\Lambda}$O.
The results are given for the J\"ulich and Nijmegen potentials and for two
approximations to the self--energy: the 
Hartree--Fock (HF) and the Hartree--Fock
plus the two--particle--one--hole diagram (HF+2P1H).
Energies are in units of MeV and $rms$ in units of fm.}
\begin{tabular}{cccccc}\hline
&\multicolumn{2}{c}{J\"ulich}&\multicolumn{2}{c}{Nijmegen}&Exp\\
&HF&HF+2P1H&HF&HF+2P1H&\\\hline
$\varepsilon_{\Lambda}$&-10.15&-11.83&-4.76&-7.38&-12.5  \cite{bando3}\\
$T$& 6.43&6.49 &4.43&5.08&\\
$rms$&2.49&2.47&3.04&2.80&\\\hline
\end{tabular}
\end{center}
\label{tab:tab2}
\end{table}

\end{document}